\documentclass[aip,amsmath,amssymb,reprint,nofootinbib 
]{revtex4-1}
\usepackage{graphicx}

\usepackage{dcolumn}
\usepackage{bm}
\usepackage[utf8]{inputenc}
\DeclareUnicodeCharacter{0301}{\'}

\usepackage[utf8]{inputenc}
\usepackage[T1]{fontenc}
\usepackage{mathptmx}
\usepackage{etoolbox}

\usepackage{xcolor}
\definecolor{myBlue}{HTML}{648FFF}
\usepackage{hyperref}
\hypersetup{
    colorlinks=true,
    linkcolor=myBlue,
    urlcolor=myBlue,
    citecolor=myBlue,
    pdftitle={Overleaf Example},
    pdfpagemode=FullScreen,
    linktoc = all
    }

\makeatletter
\def\@email#1#2{%
 \endgroup
 \patchcmd{\titleblock@produce}
  {\frontmatter@RRAPformat}
  {\frontmatter@RRAPformat{\produce@RRAP{*#1\href{mailto:#2}{#2}}}\frontmatter@RRAPformat}
  {}{}
}%
\makeatother
\begin{document}

\preprint{AIP/123-QED}

\title[]{Design and optimization of an AZO-based plasmonic metasurface-driven optical solar reflector for thermal management}
\author{L. Viseur}
 \email{lucas.viseur@unamur.be, michael.lobet@unamur.be}
\author{A. Mayer}
\author{L. Henrard}%
\author{M. Lobet}
\affiliation{ 
University of Namur, Department of Physics and Namur Institute of Structured Material
}%

\date{\today}

\begin{abstract}
\textbf{ABSTRACT}\protect\\ 
Plasmonic metasurface-driven Optical Solar Reflectors (m-OSRs) offer a promising route towards lightweight and high-performance thermal management. By exploiting subwavelength structuring and intrinsic material losses, such systems enable tailored absorptance spectrum across the solar and thermal infrared domains, respectively. Here, a plasmonic m-OSR composed of an aluminum back-reflector, a silicon dioxide dielectric spacer, and a nanostructured aluminum-doped zinc oxide (AZO) layer is investigated. The optical response of the structure is governed by the interplay between reflection, localized surface plasmon resonances and Fabry–Pérot cavity effects, leading to efficient spectral selectivity. An optimization performed with a multi-objective genetic algorithm yields a low solar absorptance of $\alpha = 0.16$ combined with a high thermal emissivity of $\varepsilon = 0.83$, providing an $\alpha/\varepsilon$ ratio of 0.19. These results highlight the potential of plasmonic meta-OSRs as ultrathin, high-performance solutions for thermal management and in particular for the next-generation advanced spacecraft.
\end{abstract}

\maketitle

\section{\label{sec:intro}INTRODUCTION}

Optical solar reflectors (OSRs) are designed to minimize solar absorptance $\alpha$ while maximizing thermal infrared emissivity $\varepsilon$, enabling passive radiative thermal management.\cite{osr_def,Marshall} Achieving low $\alpha$ together with high $\varepsilon$ requires strong spectral selectivity across two widely separated wavelength bands (\autoref{fig:1}\textcolor{myBlue}{a}). Such spectral control is particularly relevant for space-based thermal regulation, \cite{nextgenscraft,osr_spacecraft1,osr_spacecraft2} where radiative exchange dominates heat transfer, or all-day radiative coolers.\cite{radiation,Ta2O5,AZO}

Conventional OSRs are based on secondary surface mirrors composed of a metallic layer deposited on glass.\cite{moser_conv,classical_osr,Marshall,Ta2O5} While such systems provide excellent thermo-optical performance ($\alpha = 0.08$–0.15 ; $\varepsilon = 0.8$–0.9), they remain relatively thick, fragile, and limited in geometric flexibility.\cite{moser_conv,classical_osr,Ta2O5} More recent approaches have explored lightweight and flexible alternatives (PMMA, fluorinated ethylene polymer),\cite{Townsend,Putz,Wang,Ta2O5} yet often at the expense of long-term stability due to UV degradation, or spectral performance.\cite{Townsend,Putz,Wang,Ta2O5}

In this context, metasurfaces, engineered interfaces composed of arrays of subwavelength elements that enable unconventional light–matter interactions and spectral tailoring,\cite{Capasso,MS_Huang,ms_def} offer a fundamentally different route to OSR.\cite{Chen,AZO,Ta2O5} By structuring materials at subwavelength scales, it is possible to tailor light–matter interactions within ultrathin architectures.\cite{Capasso,MS_Huang} Metasurfaces applications include flat optics and metalenses,\cite{flat_optics,metalenses} holography,\cite{holography} optical cloaking and stealth\cite{cloak,Chen} and many more.\cite{ms_review} Several strategies have emerged for metasurface-driven OSRs, such as interference-based dielectric multilayers exploiting Fabry–Pérot effects to shape emissivity spectra,\cite{Rephaeli,Ta2O5}, high-index dielectric resonators relying on Mie or phonon-polariton resonances,\cite{Shen,Ding,Murai} and finally plasmonic metasurfaces,\cite{AZO} exploiting material losses and localized surface plasmon resonances (LSPR).\cite{lspr_book,maier,LSPR} Interference-based OSR offers excellent solar reflectance and low fabrication cost. However, the required number of layers increases the complexity and mass, and emissivity remains moderate compared to loss-driven approaches.\cite{Rephaeli,Ta2O5} Dielectric resonators OSR enables angularly robust and low-loss designs but often yields narrowband emission peaks, making it better suited to Earth-based radiative cooling where the atmospheric window plays a role.\cite{Shen,Ding,Murai} In space, where the full infrared range is relevant, broadband solutions are preferred.\cite{nextgenscraft,ECSS} Plasmonic architectures combining a metallic back-reflector, a dielectric spacer, and a structured conductive layer provide multiple degrees of freedom through geometry, thickness, and carrier concentration control.\cite{AZO} However, the interplay between reflection, cavity resonances, plasmonic modes, and intrinsic material dispersion remains not sufficiently clarified. In particular, systematic multi-parameter optimization grounded in physical insight is still scarce. Its main drawbacks are higher fabrication cost and increased solar absorption due to losses.\vspace{0.1cm}\newline
\indent In this work, we investigate a plasmonic m-OSR based on a trilayer architecture of aluminum (Al), silicium dioxide (SiO$_2$) and aluminum-doped zinc oxide (AZO). We first analyze the role of each layer and identify the dominant physical mechanisms governing spectral selectivity. Building on this understanding to limit the search domain, a multi-objective genetic algorithm \cite{Mayer_SPIE2018,algo1,algo2} is employed to optimize the design with respect to solar absorptance $\alpha$ and thermal emissivity $\varepsilon$. We demonstrate that by combining subwavelength structuration with lossy conductive materials, plasmonic m-OSRs enable the simultaneous excitation of multiple resonant mechanisms within an ultrathin architecture. Localized surface plasmon resonances, Fabry-Pérot modes, and back-reflection can be exploited to tailor the design spectral response, resulting in high emissivity together with low solar absorptance within a reduced number of layers. The resulting compact and lightweight structure achieves high emissivity with low solar absorption while preserving angular robustness, highlighting its performance as an efficient optical solar reflector.

\section{MATERIALS AND METHODS}

\begin{figure*}
\includegraphics[width=1\textwidth,trim={0cm 6cm 0cm 4cm},clip]{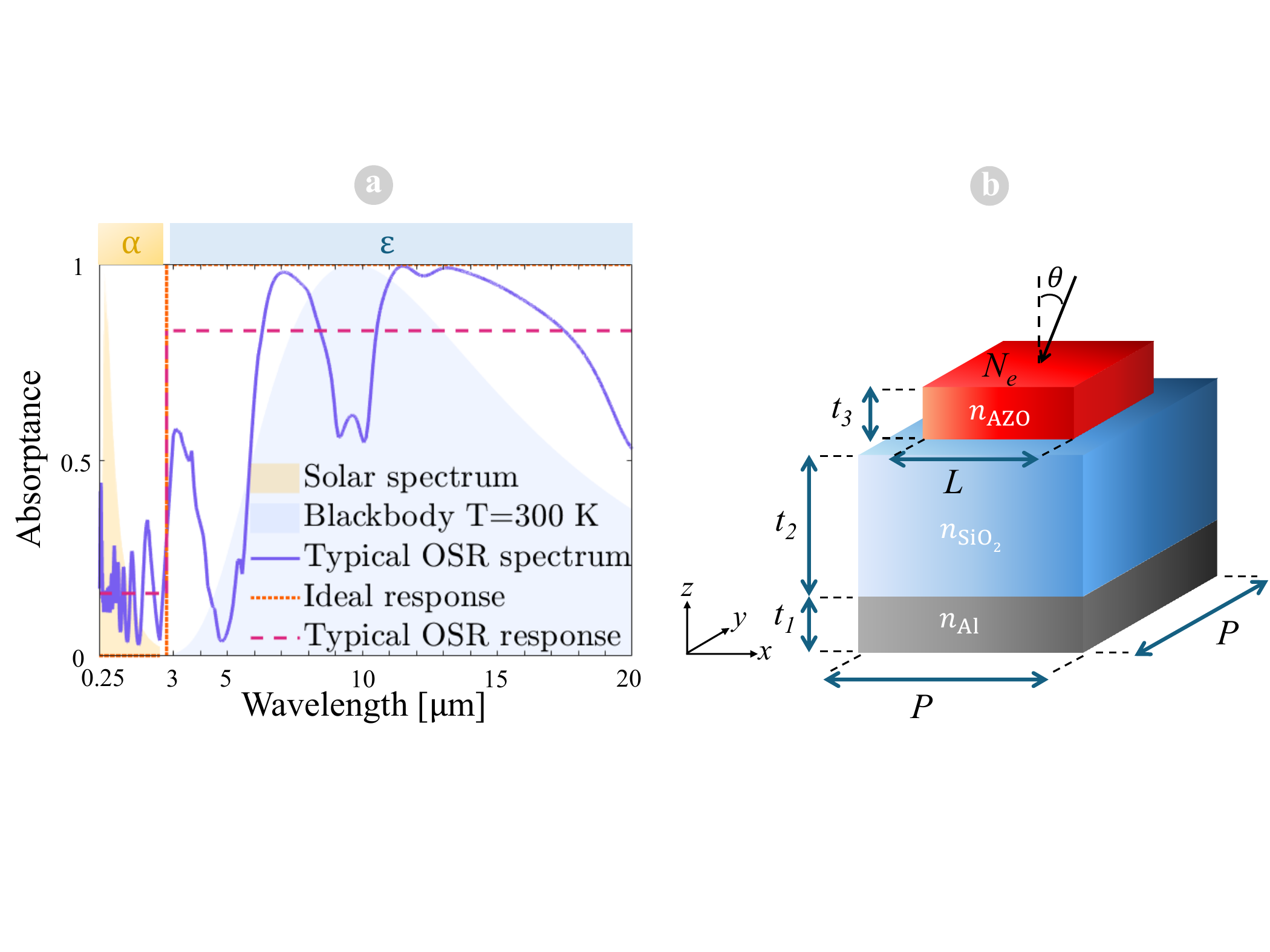}
\caption{\label{fig:1}(a) OSR ideal response (orange dashed line) in contrast with typical averaged OSR response (pink dashed line) and spectrum (violet curve) in both solar (yellow shadowed area, 0.25 to 2.5 microns) and thermal infrared (blue shadowed area, 3 to 20 microns) ranges. (b) Investigated design and associated parameters of a plasmonic metasurface-driven optical solar reflector made of a trilayer Al/SiO$_2$/AZO: layer thicknesses $t_1$, $t_2$, and $t_3$, the unit cell period $P$, the lateral dimension $L$ of the AZO square and AZO carrier concentration $N_e$. $n_{\text{AZO}}$, $n_{\text{SiO$_2$}}$ and $n_{\text{Al}}$ are the refractive indices of AZO, SiO$_2$ and Al respectively and $\theta$ is the angle of incidence.}
\end{figure*}

Performance is quantified through the following figures of merit according to the European Cooperation for Space Standardization conventions:\cite{ECSS} the spectrally integrated solar absorptance
 
\begin{equation}
\alpha=\frac{\int_{\lambda_1}^{\lambda_2}{A\left(\lambda\right)B_s\left(\lambda\right)d\lambda}}{\int_{\lambda_1}^{\lambda_2}{B_s\left(\lambda\right)d\lambda}},\label{eq:alpha}
\end{equation}
integrated between $\lambda_1=0.3\ \mu$m and $\lambda_2=2.5\ \mu$m with the ASTM E-490 \cite{ECSS} solar irradiance $B_s\left(\lambda\right)$ and the spectral absorptance $A\left(\lambda\right)$, and the thermal emissivity 
\begin{equation}                                   \varepsilon=\frac{\int_{\lambda_1}^{\lambda_2}A\left(\lambda\right)B\left(\lambda,T\right)d\lambda}{\int_{\lambda_1}^{\lambda_2}B\left(\lambda,T\right)d\lambda},\label{eq:epsilon}
\end{equation}
integrated between $\lambda_1=3\ \mu$m and $\lambda_2=20\ \mu$m with the blackbody emission $B\left(\lambda,T\right)$ taken at $T=300$ K.

An ideal OSR's performance is depicted by the orange dashed curve in \autoref{fig:1}\textcolor{myBlue}{a}. This fictitious OSR totally reflects the solar spectrum (0.25 to 2.5 microns), hence absorptance $A=0$, and behaves like a perfect blackbody in the thermal infrared (3 to 20 microns), hence $A=1$. On the other hand, a realistic OSR exhibit an absorptance spectrum similar to the violet continuous curve, as illustrated in \autoref{fig:1}\textcolor{myBlue}{a}, resulting in a small absorption in the solar spectrum and a deviation from the blackbody behavior in the thermal infrared. The overall performance of this realistic OSR is depicted by the dashed pink line in \autoref{fig:1}\textcolor{myBlue}{a}, where the averaged value of $\alpha$ ($\varepsilon$) is represented throughout the solar spectrum (thermal infrared) by an horizontal line according to \autoref{eq:alpha} (\autoref{eq:epsilon}). The subsequent optimization process consists in converging this realistic response towards the ideal one by tuning the different parameters of the design.

Our plasmonic m-OSR consists of a periodic trilayer Al/SiO$_2$/AZO architecture disposed in a square array, for which the unit cell is illustrated in \autoref{fig:1}\textcolor{myBlue}{b}. The plasmonic AZO layer is nanostructured, while the dielectric spacer and the aluminum back-reflector form a resonant cavity analogous to a Salisbury screen.\cite{Salisbury} Six parameters are left as tunable variables for the optimization, including the layer thicknesses $t_1$, $t_2$, and $t_3$ of Al, SiO$_2$ and AZO, the unit cell period $P$, and the lateral dimension $L$ of the AZO square. The AZO free carrier concentration $N_e$ is also considered as a tunable parameter. All other parameters, as described below, are fixed. The permittivity of AZO is described using a Drude–Lorentz model with parameters derived from Sun \textit{et al.}:\cite{AZO}

\begin{equation}
    \varepsilon(\omega) = \varepsilon_{\infty} - \frac{\omega_p^2}{\omega^2 + i\omega\Gamma} + \frac{f_1\omega_1^2}{\omega_1^2-\omega^2+i2\omega\Gamma_1},
\end{equation}

\noindent with the plasma frequency $\omega_p$, the background permittivity $\varepsilon_{\infty}$ and the damping constant $\Gamma$ of the Drude model. Parameters $f_1$, $\omega_1$ and $\Gamma_1$, are respectively the amplitude, spectral position and damping constant of the Lorentz oscillator accounting for the AZO ultraviolet absorptive peak around 0.3 $\mu$m. The plasma frequency further depends on the carrier concentration $N_e$, the elementary charge $e$, the effective electron mass $m^*$ and the vacuum permittivity $\varepsilon_0$ through

\begin{equation}
    \omega_p^2= \frac{N_e e^2}{m^*\varepsilon_0}.
\end{equation}

\noindent The permittivity of SiO$_2$ is retrieved from Franta \textit{et al.}\cite{Franta} while Al is described by a Drude model provided in Rakic \textit{et al.}\cite{Rakic}. The fixed parameters of the associated models and the permittivities of all materials used in this study are provided in \textcolor{myBlue}{Appendix} \ref{appendixes}. Finally, the filling fraction is defined by $L/P$, whereas the aspect ratio is expressed by $t_3/L$.

\begin{figure*}
\includegraphics[width=1\textwidth,trim={0cm 7.5cm 0cm 2.5cm},clip]{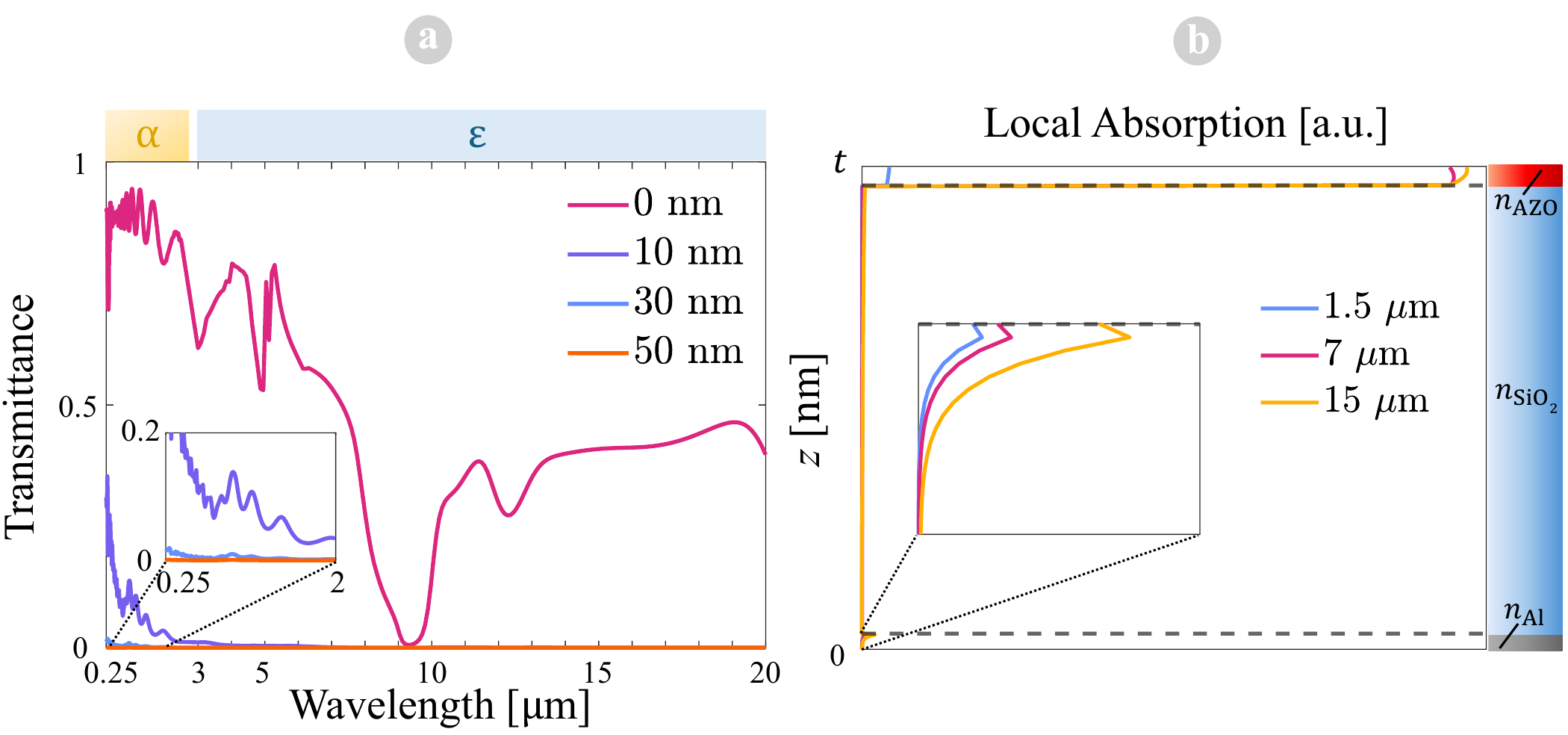}
\caption{\label{fig:2}(a) Transmittance spectra of the plasmonic m-OSR design for increasing values of Al thickness $t_1$ from 0 to 50 nm in both solar and thermal infrared ranges. (b) Qualitative local absorption in the three constitutive layers of total thickness $t$ at different wavelengths: 1.5, 7 and 15 microns.}
\end{figure*}

\begin{figure*}
\includegraphics[width=1\textwidth,trim={0cm 7.5cm 0cm 2.5cm},clip]{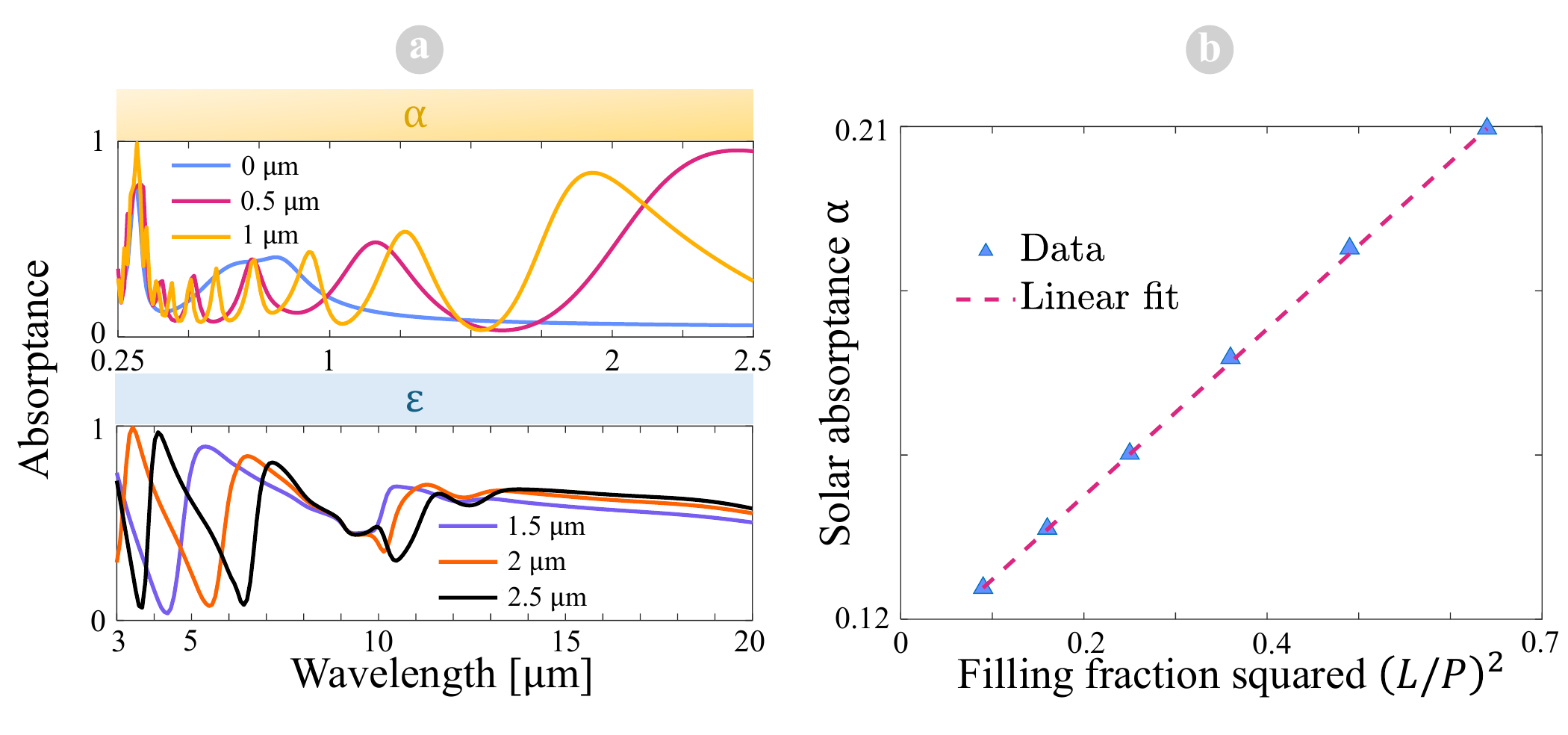}
\par
\includegraphics[width=1\textwidth,trim={0.5cm 6.5cm 1cm 4.5cm},clip]{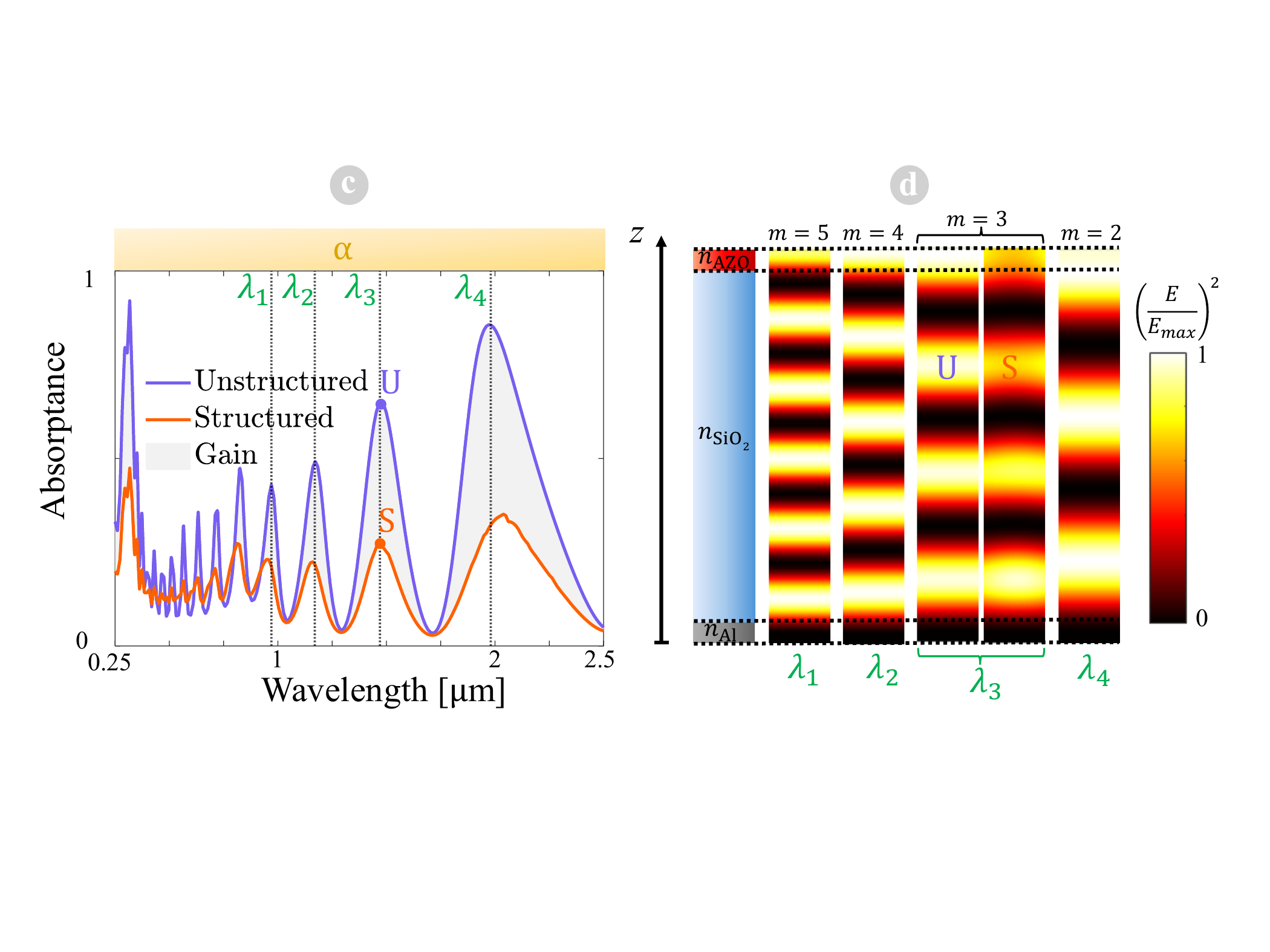}
\caption{\label{fig:3}(a) Influence of the thickness $t_2$ of SiO$_2$ on the absorptance spectrum for the unstructured design in both solar and thermal infrared ranges. (b) Relation between the computed solar absorption figure of merit $\alpha$ and the square of the filling fraction of the design. (c) Comparison of the absorptance spectrum of the unstructured and structured (filling fraction of 0.56) design in the solar spectrum range. (d) Cross-sectional field maps of the design at specific wavelengths marked in (c).}
\end{figure*}

Each layer is first investigated separately, and the occurring phenomena are studied at normal incidence unless stated otherwise. The influence of the different parameters of the layer on the overall performance is then assessed. Finally, as interplay between different phenomena or parameters is inevitable, their interactions are looked upon. Physical understanding of the influence of the parameters allows us to refine the optimization range of each parameter to subsequently exploit a genetic algorithm and perform a multiobjective optimization. \cite{algo1,algo2,Mayer_SPIE2018} The optical response is computed using a rigorous coupled-wave analysis (RCWA) solver developed in-house.\cite{RCWA} Angular robustness of the optimized design is ensured and assessed \textit{a posteriori}, triggering a potential feedback loop.

A genetic algorithm is a meta-heuristic used to determine solutions to an optimization problem. Genetic algorithms in particular take inspiration from the process of natural selection, employing mechanisms such as selection, crossover or mutation inherent to biology.\cite{algo1,Eiben_2007,Haupt_2007} For our plasmonic m-OSR, we want to minimize $\alpha$ while maximizing $\varepsilon$. Objective functions $f_1 = 1-\alpha$ and $f_2 = \varepsilon$ are therefore implemented to work instead on two functions to be maximized. The particularity of the multi-objective genetic algorithm is in emphasizing non-dominated solutions, solutions that cannot improve one of the figures of merit ($\alpha$ or $\varepsilon$) without deteriorating the other. The resulting set of non-dominated solutions is called the Pareto front.\cite{algo2}

\newpage
\section{RESULTS AND DISCUSSION}
\subsection{Impact of the parameters}

\textbf{Back-reflector layer}\protect\\

The bottom layer consists of an aluminum back-reflector, ensuring high reflectivity in the solar spectrum and suppressing transmission over the full spectral range of interest. The thickness must therefore exceed the characteristic skin depth $\delta$ of aluminum, while remaining compatible with mass constraints. Following its definition at a specific wavelength of light in vacuum $\lambda_0$,\cite{maier}

\begin{equation}
    \delta = \frac{\lambda_0}{4\pi k},
\end{equation}

\noindent it can be computed using the imaginary part of the refractive index of aluminum $k$.\cite{Rakic} Resulting values span from 6 to 11 nm throughout the spectrum; a thickness of $5\delta \approx 55$ nm thus decreases the transmission to less than 1\% of its initial value.
\autoref{fig:2}\textcolor{myBlue}{a} shows the spectral transmittance for Al thicknesses ranging from 0 to 50 nm to confirm this trend. Increasing the thickness rapidly suppresses transmission across the 0.25–20 $\mu$m range. For thicknesses approaching 50 nm, transmission becomes negligible, consistent with the computed skin depth of aluminum in both spectral domains.\cite{maier} To further elucidate the field distribution, a quantity proportional to local absorption and defined by $\textit{Im}(\varepsilon)|E|^2$\cite{local_ab_brenner} spanning each layer is evaluated at representative wavelengths through the solar and thermal infrared ranges ($\lambda=1.5\mu$m, $\lambda=7 \mu$m and $\lambda=15 \mu$m), as shown in \autoref{fig:2}\textcolor{myBlue}{b}. The SiO$_2$ spacer remains nearly transparent throughout the range, in agreement with the low imaginary part of its permittivity $Im(\varepsilon)$ \cite{Franta} (see \textcolor{myBlue}{Appendix} \ref{appendixes}). In contrast, the AZO layer exhibits significant absorption, increasing toward longer wavelengths, as described by its Drude–Lorentz dispersion \cite{AZO} (see \textcolor{myBlue}{Appendix} \ref{appendixes}). Minor deviations from the expected exponential decay characteristic of metallic skin-depth-limited penetration\cite{maier} at the dielectric interface arise from numerical discretization.

A thickness of 60 nm is therefore selected to ensure complete transmission suppression with a safety margin. This choice effectively fixes the back-reflector response and reduces the dimensionality of the optimization problem to five tunable parameters.

A "simple" mirror suppressing transmission humbly constitutes the first physical mechanism enabling thermal control in the plasmonic m-OSR.
\protect\\

\newpage
\textbf{Dielectric spacer}\protect\\

To decouple the effect of the spacer thickness from other geometrical parameters, the SiO$_2$ thickness $t_2$ is first analyzed for unstructured "bulk" AZO ($L=P$). \autoref{fig:3}\textcolor{myBlue}{a} presents the corresponding spectra, with solar and thermal infrared domains shown separately for clarity.

\begin{figure*}
\includegraphics[width=1\textwidth,trim={0cm 7.5cm 0cm 5cm},clip]{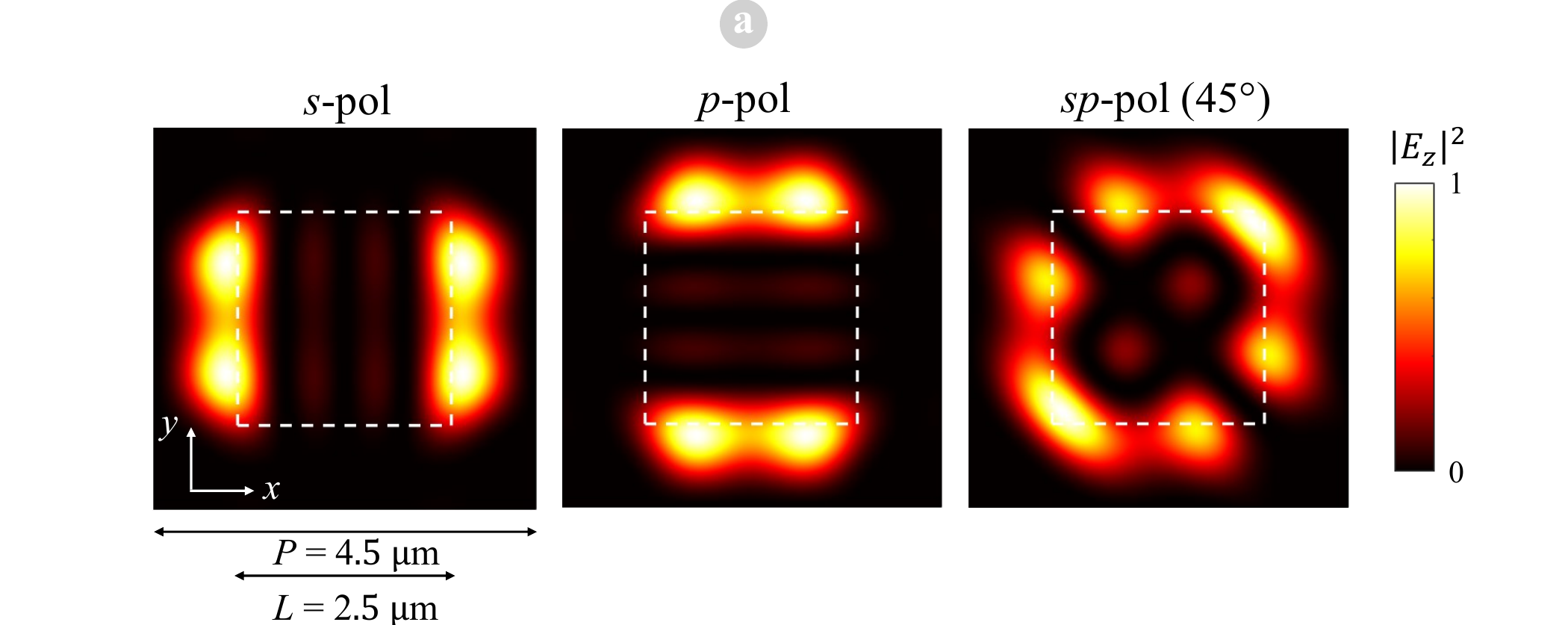}
\par
\includegraphics[width=1\textwidth,trim={0cm 7cm 0cm 4.5cm},clip]{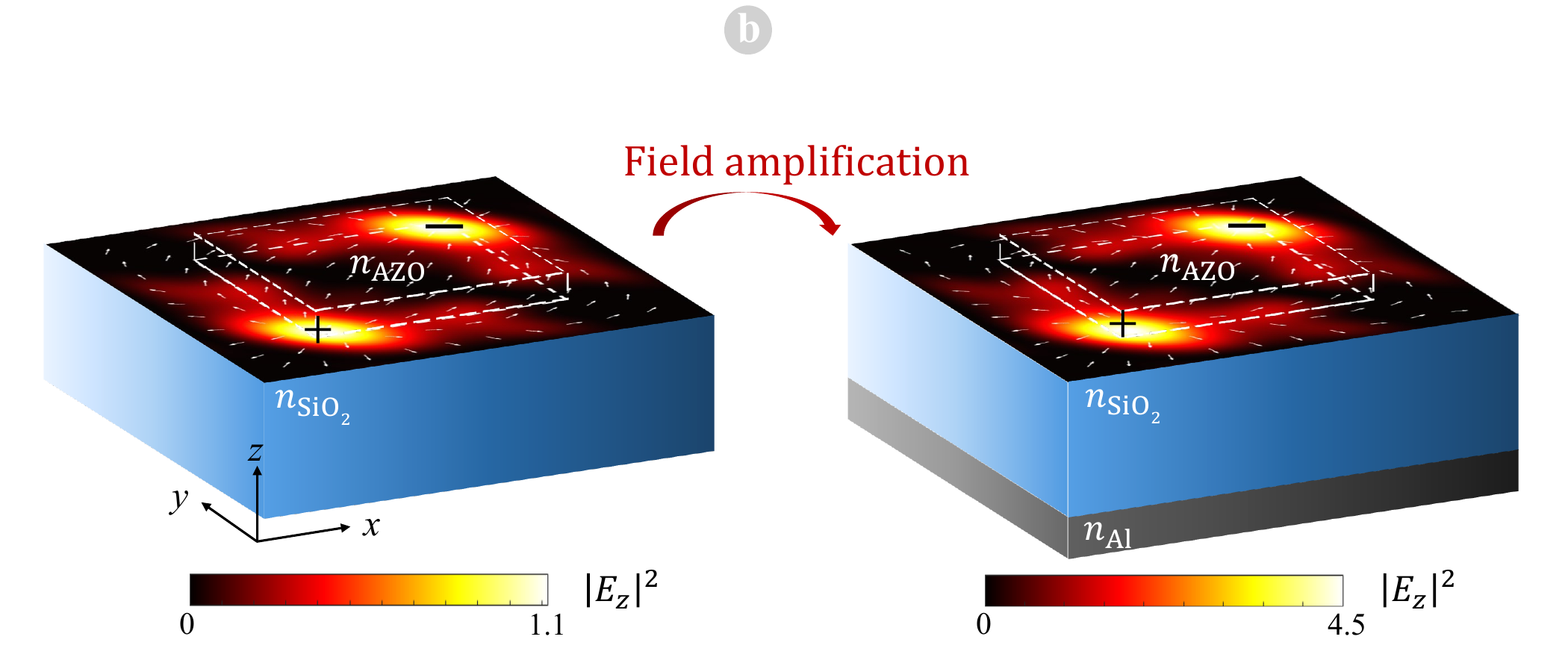}
\par
\includegraphics[width=1\textwidth,trim={0cm 6.5cm 0cm 4.5cm},clip]{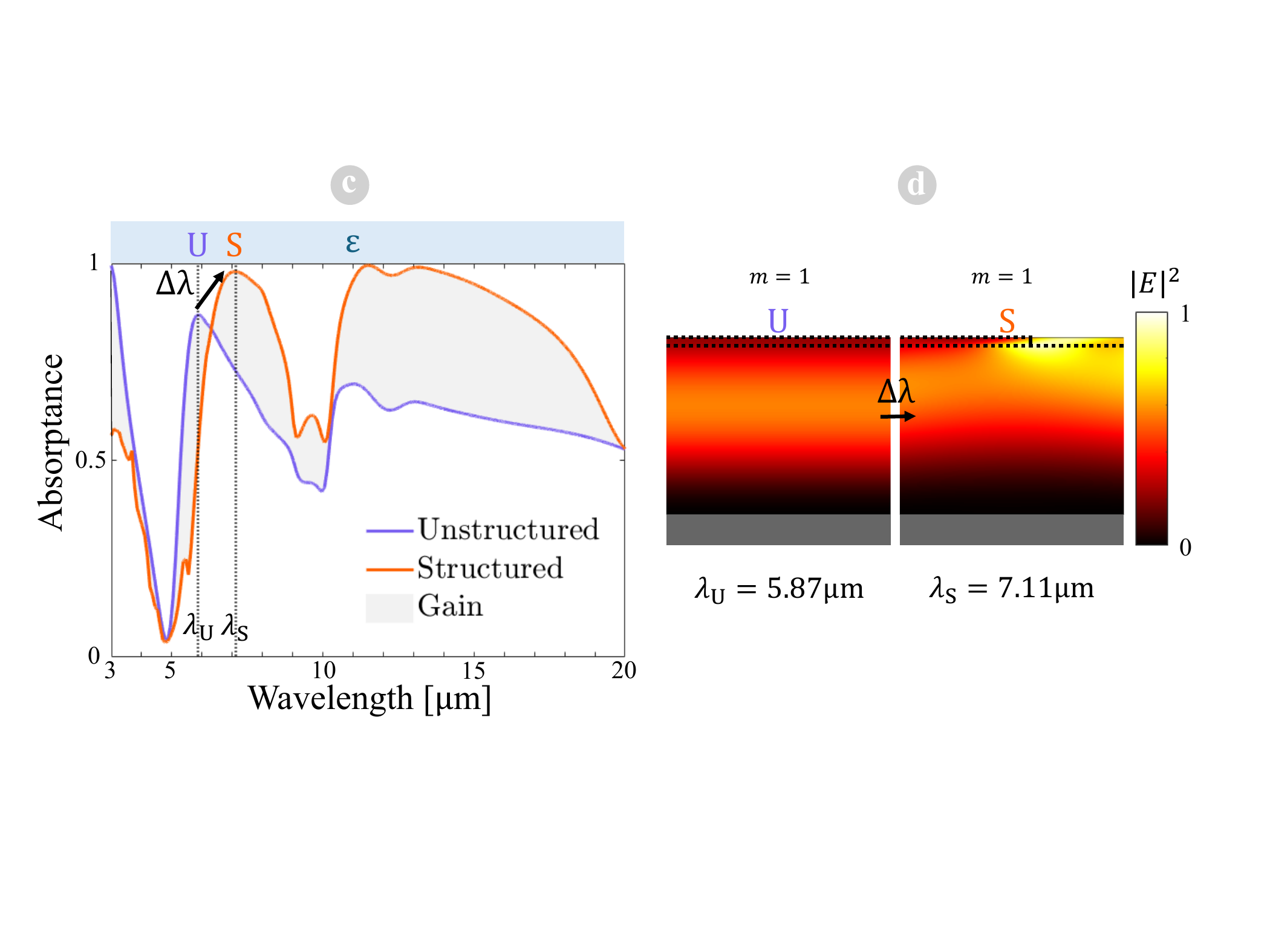}
\caption{\label{fig:4}(a) LSPR field maps of the periodical structured AZO layer only for different polarizations. (b) Field amplification of a bilayer AZO/SiO$_2$ design by addition of a third Al layer. (c) Comparison of the absorptance spectrum of the unstructured and structured (filling fraction of 0.56) design in the thermal infrared range. (d) Comparison for U ($\lambda = 5.87$ $\mu$m) and S ($\lambda = 7.11$ $\mu$m) of the associated cross-sectional field maps across $\Delta\lambda$ specified in (c).}
\end{figure*}

In the absence of a dielectric spacer, $t_2=0$ nm, the response in the solar range is dominated by the aluminum back-reflector, resulting in high reflectance and low absorptance. The absorption peak around 0.3 $\mu$m originates from the AZO UV absorptive peak described by the Lorentz oscillator,\cite{AZO} as illustrated in \textcolor{myBlue}{Appendix} \ref{appendixes}.
 
As soon as a finite spacer thickness is introduced, Fabry-Pérot modes emerge and progressively transition to longer wavelengths while increasing $t_2$, from the solar spectrum toward the thermal infrared (\autoref{fig:3}\textcolor{myBlue}{a}). To describe this transition, the shift is shown in the solar spectrum for $t_2$ values of 500 and 1000 nm and in the thermal infrared for greater values of 1500, 2000, and 2500 nm. This behavior arises from the cavity formed by the SiO$_2$ layer between the metallic back-reflector (Al) and the partially reflecting AZO layer.

To confirm this interpretation, cross-sectional field maps are evaluated at wavelengths corresponding to absorptance maxima in \autoref{fig:3}\textcolor{myBlue}{c-d} for the unstructured design. The field distributions reveal standing-wave patterns inherent to Fabry-Pérot modes confined within the SiO$_2$ spacer, with modal orders decreasing from $m=5$ at $\lambda_1 = 0.97$ $\mu$m to $m=2$ at $\lambda_4=1.98$ $\mu$m (\autoref{fig:3}\textcolor{myBlue}{d}).

The Fabry-Pérot modes are also influenced by the AZO layer structuration. This effect is now investigated by first considering an effective medium approach based on the filling fraction. In the solar domain, the absorptance $\alpha$ exhibits a linear dependence on the AZO filling fraction squared (\autoref{fig:3}\textcolor{myBlue}{b}), independently of the remaining degrees of freedom. This linear scaling indicates that structuration primarily modulates the resonance amplitude through the effective filling fraction, without significantly altering their spectral positions. This observation is confirmed in \autoref{fig:3}\textcolor{myBlue}{c}. Since AZO is an absorptive material in the solar spectrum (see \textcolor{myBlue}{Appendix} \ref{appendixes}), reduction in AZO coverage leading to a proportional decrease in $\alpha$ is expected.

In \autoref{fig:3}\textcolor{myBlue}{d}, field maps at $\lambda_3=1.47$ $\mu$m further support this interpretation, showing similar modal distributions for structured (S) and unstructured (U) configurations, with reduced field intensity in the former case. However, for longer wavelengths such as $\lambda_4=1.98$ $\mu$m, a slight spectral shift becomes apparent (\autoref{fig:3}\textcolor{myBlue}{c}) between the unstructured and structured cases, indicating that structuration progressively modifies the effective cavity response as the wavelength increases. This regime will be addressed in the following section.

Fabry–Pérot resonances within the dielectric spacer thus constitute the second key physical mechanism governing thermal control in the plasmonic m-OSR.\protect\\

\textbf{Plasmonic layer}\protect\\

The role of AZO structuration is first examined in the thermal infrared for an isolated periodic AZO layer design ($t_1=t_2=0$ nm, neither the dielectric nor the back-reflector is present). While structuration was shown to primarily modulate resonance amplitude in the solar range without introducing new mechanisms (\autoref{fig:3}\textcolor{myBlue}{c-d}), a different behavior emerges in the thermal infrared. Patterning enables the excitation of localized surface plasmon resonances (LSPR),\cite{maier,lspr_book} leading to strong near-field confinement and enhanced absorption.\cite{LSPR,lspr1,lspr2,lspr3} Capitalizing on this mechanism is particularly relevant in the thermal infrared, where high absorptance is desired.\cite{algo1,Mika}

\autoref{fig:4}\textcolor{myBlue}{a} shows the LSPR response of the periodic AZO square at the resonance wavelength ($\lambda=9.1$ $\mu$m) for different polarizations. Edge modes are observed at the AZO surfaces.\cite{edgemodes} At normal incidence, polarization modifies the field distribution but does not affect the spectral response (not shown); subsequent calculations are performed for a 45° polarization angle without loss of generality. This serves a dual purpose by optimizing near-field visualization by displaying both orthogonal modes of the metasurface and acting as an efficient computational proxy for unpolarized ambient light, such as sunlight, which exhibits no preferred polarization axis. LSPR at the patterned interface constitute the third key physical mechanism contributing to thermal control in the m-OSR.

The impact of integrating the spacer and the back-reflector is shown in \autoref{fig:4}\textcolor{myBlue}{b}. Adding the SiO$_2$ layer preserves comparable field amplitude (from 1 to 1.1, normalized), while the introduction of the metallic back-reflector produces a pronounced field enhancement, yielding a significant increase in absorptance in the thermal infrared (from 1.1 to 4.5, normalized), also described by the structuration gain in \autoref{fig:4}\textcolor{myBlue}{c}. The presence of the mirror indeed suppresses transmission, leading to the possibility to harvest energy that was lost without the reflector. This illustrates a clear coupling between reflection and LSPR.

\autoref{fig:4}\textcolor{myBlue}{c} presents the fully integrated m-OSR response. By contrast to the solar range (\autoref{fig:3}\textcolor{myBlue}{c}), structuration impacts both the amplitude and the spectral position of the absorption peaks. In the absence of structuration, the response remains governed by Fabry–Pérot resonances. A filling fraction argument would suggest that reducing the AZO coverage would decrease the overall emittance $\varepsilon$, analogously to the behavior observed for the solar absorptance $\alpha$ in the dielectric spacer section (\autoref{fig:3}\textcolor{myBlue}{b}). However, \autoref{fig:4}\textcolor{myBlue}{c} reveals the opposite trend: structuration induces a substantial increase in emittance. This enhancement results from the interplay between LSPR and Fabry–Pérot cavity modes. Cross-sectional field maps (\autoref{fig:4}\textcolor{myBlue}{d}) computed at the absorption peaks identified for the unstructured (U) and structured (S) configurations, associated with a wavelength shift $\Delta\lambda$, confirm the coexistence of both mechanisms. This shift can be understood by the change in interference conditions given by the structuration. A standing-wave pattern (modal order $m=1$) persists within the SiO$_2$ cavity, while pronounced localized field enhancement appears at the AZO edges. Structuration therefore introduces LSPR without suppressing the underlying Fabry–Pérot resonance, albeit reshaping the coupled modal response.

Finally, a fourth effect is identified in the 8–11 $\mu$m range (\autoref{fig:4}\textcolor{myBlue}{c}), corresponding to the Reststrahlen band of SiO$_2$.\cite{Reststrahlen,Ding} In this spectral region, the real part of the permittivity of SiO$_2$ becomes negative (\textcolor{myBlue}{Appendix} \ref{appendixes}), leading to metal-like behavior and inducing strong reflectivity. This phonon–polariton resonance limits absorption and degrades the m-OSR performance. This intrinsic material response must therefore be accounted for during optimization.

Four mechanisms are thus intertwined in the complete m-OSR: reflection from the metallic back-reflector, Fabry–Pérot resonances within the dielectric spacer, localized surface plasmon resonances in the patterned AZO layer, and Reststrahlen band of SiO$_2$. Building on the understanding of these four coupled phenomena, multi-objective optimization is performed in the next section.

\subsection{Optimization}

\begin{figure*}
\includegraphics[width=1\textwidth,trim={0cm 9cm 0cm 4cm},clip]{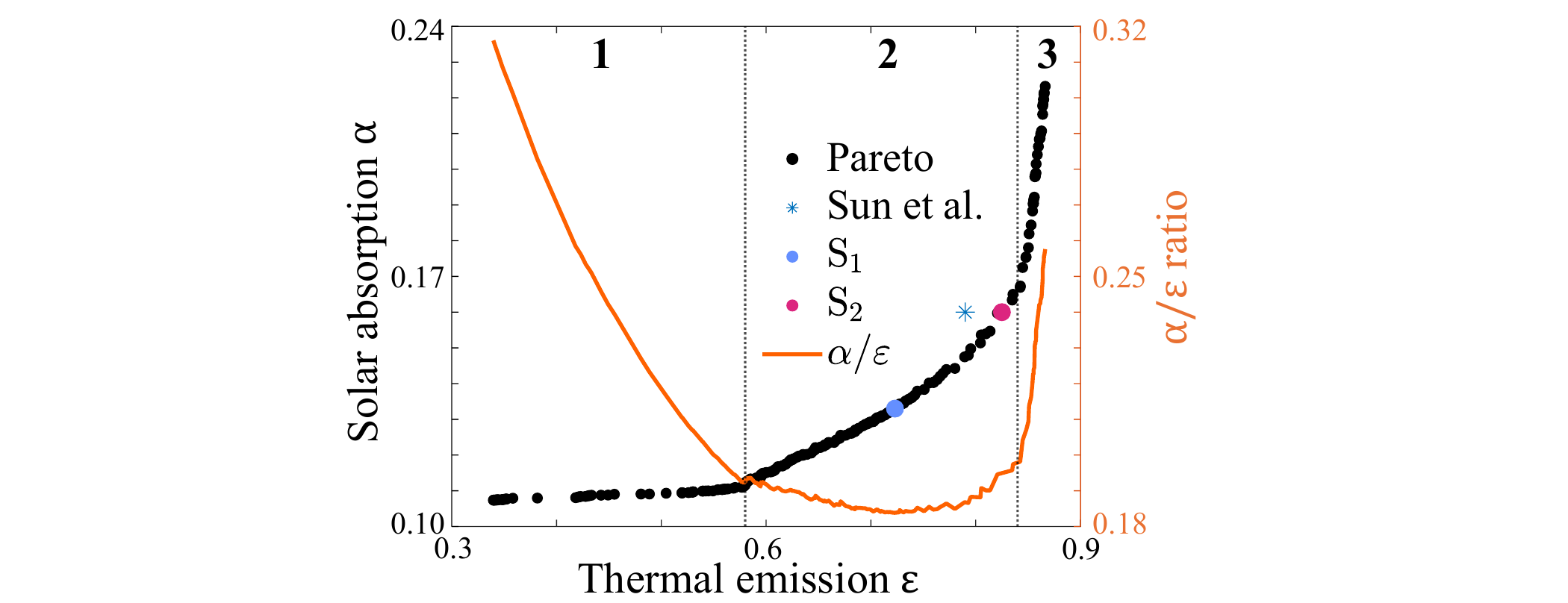}
\caption{\label{fig:5}Pareto front obtained through the optimization process and associated solution from Sun \textit{et al.} \cite{AZO} and optimized solutions $S_1$ ($\alpha=0.13$; $\varepsilon=0.72$) and $S_2$ ($\alpha=0.16$; $\varepsilon=0.83$) in the figure of merits space.}
\end{figure*}
\begin{figure*}
\includegraphics[width=1\textwidth,trim={0cm 7cm 0cm 1cm},clip]{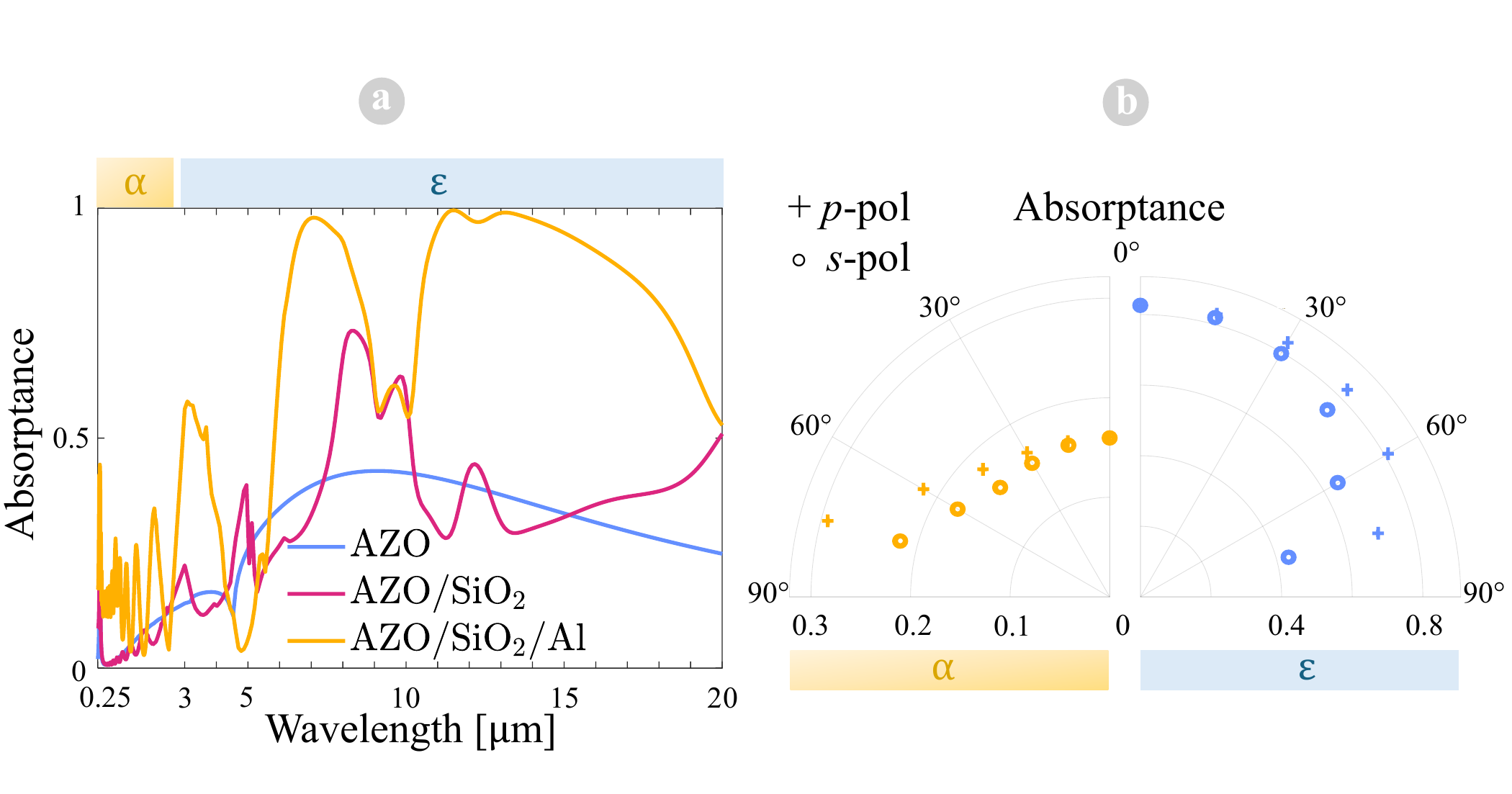}
\caption{\label{fig:6}(a) Optimized absorptance spectrum of the m-OSR design (solution $S_2$) and layer by layer contribution to the overall performance. (b) Angular dependence of the design for both $s$ and $p$ polarizations.}
\end{figure*}

The design to optimize initially involved six parameters. By fixing the aluminum thickness to $t_1=60$ nm to suppress transmission, we reduce the search space to five variables: $L$, $P$, $t_2$, $t_3$, and $N_e$ (\autoref{fig:1}\textcolor{myBlue}{b}).

The lateral feature size $L$ is explored between 250 and 3000 nm (1 nm step), while the period $P$ is varied such that the longitudinal gap defined by ($(P-L)/2$) ranges from 50 to 1000 nm (1 nm step). A finite gap is required to sustain LSPR, whereas excessively large separations reduce the effective filling fraction and limit emissivity enhancement.

The SiO$_2$ spacer thickness $t_2$ is varied between 1200 and 3000 nm (1 nm step), ensuring Fabry–Pérot resonances within the spectral region of interest.

For AZO, the thickness $t_3$ is optimized in the range 5 to 250 nm (1 nm step) and the carrier concentration $N_e$ between 0 and $8\cdot 10^{20}$ cm$^{-3}$ (step of $0.5\cdot 10^{20}$ cm$^{-3}$), in agreement with experimentally achievable values.\cite{AZOdoping} These parameters introduce intrinsic trade-offs: increasing either $t_3$ or $N_e$ enhances thermal emissivity but simultaneously increases solar absorptance. Moreover, higher values of the AZO thickness and/or carrier concentration render the plasmonic layer progressively opaque, leading to saturation of emissivity gains due to reduced field penetration into the cavity.
 
The resulting multi-objective optimization yields the Pareto front shown in \autoref{fig:5}, which delineates the set of non-dominated solutions in the ($\alpha$, $\varepsilon$) space. The design proposed by Sun \textit{et al.} \cite{AZO} lies strictly inside the Pareto front, indicating that simultaneous improvement of both figures of merit is achievable within the present parameter space.

Two representative solutions are selected. $S_1$ minimizes the ratio $\alpha/\varepsilon$, whereas $S_2$ improves both $\alpha$ and $\varepsilon$ relative to Sun \textit{et al.}\cite{AZO} While $S_1$ exhibits a lower solar absorptance ($\alpha=0.13$, $\varepsilon=0.72$), $S_2$ provides a substantially higher emissivity ($\alpha=0.16$, $\varepsilon=0.83$) with only a moderate increase in $\alpha$ (+0.03). Owing to this balanced performance, $S_2$ is retained for subsequent analysis. The corresponding optimized parameters are summarized in \autoref{tab:param}, the main difference being in a larger and thicker AZO layer in a smaller unit cell to reach higher permittivity values for $S_2$.

\begin{table}
\caption{Performance and associated parameters values for solutions $S_1$ and $S_2$ of the plasmonic metasurface-driven Optical Solar Reflector.}
\begin{ruledtabular}
\begin{tabular}{lcr}
$S_1$&$\alpha=0.13$&$\varepsilon=0.72$\\
\hline
Parameter&Optimized value&Units\\
\hline
$P$ & 6159&nm\\
$L$ & 2301&nm\\
$t_2$ & 2147&nm\\
$t_3$ & 13&nm\\
$N_e$ & $8 \cdot 10^{20}$ &cm$^{-3}$\\
\end{tabular}
\end{ruledtabular}
\begin{ruledtabular}
\begin{tabular}{lcr}
$S_2$&$\alpha=0.16$&$\varepsilon=0.83$\\
\hline
Parameter&Optimized value&Units\\
\hline
$P$ & 4522&nm\\
$L$ & 2526&nm\\
$t_2$ & 1713&nm\\
$t_3$ & 73&nm\\
$N_e$ & $6.5 \cdot 10^{20}$ &cm$^{-3}$\\
\end{tabular}
\end{ruledtabular}
\label{tab:param}
\end{table}

The optimized absorptance spectrum and layer-resolved contributions are presented in \autoref{fig:6}\textcolor{myBlue}{a}. For the periodic AZO square design, only the LSPR contribution is observed in the thermal range with a resonant peak at $\lambda = 9.1$ $\mu$m, while solar absorption remains significant. Introducing the SiO$_2$ spacer activates Fabry–Pérot resonances, yielding multiple absorption peaks. The complete stack, including the aluminum back-reflector, suppresses transmission and enhances field confinement, resulting in increased absorptance within the active layers. The Reststrahlen band of SiO$_2$ remains visible around 10 $\mu$m, limiting performance in this spectral window.

Although optimized for normal incidence, angular robustness is evaluated in \autoref{fig:6}\textcolor{myBlue}{b} for both $s$ and $p$ polarizations. Performance remains stable up to approximately 45°, beyond which both solar reflection and thermal emissivity degrade. Since the incident radiative flux decreases with increasing angle according to Lambert's cosine law,\cite{Lambert} this range is compatible with practical operating conditions.

\clearpage
\section{CONCLUSIONS}

This work demonstrated and optimized a plasmonic metasurface-driven optical solar reflector based on a three-layer Al/SiO$_2$/AZO architecture. By identifying and taking advantage of four coupled physical mechanisms—metallic reflection, Fabry–Pérot cavity resonances, localized surface plasmon resonances, and the Reststrahlen band of the dielectric spacer—a multi-objective genetic algorithm was employed to tailor the spectral response.

The optimized design achieves a solar absorptance of $\alpha=0.16$ and a thermal emissivity of $\varepsilon=0.83$, while maintaining angular robustness up to 45°. The resulting structure remains compact and lightweight, with a total thickness below 2 $\mu$m and a lateral periodicity below 4.5 $\mu$m, providing enhanced flexibility with respect to the incumbent classical OSR, at the slight expense to performance \cite{Ta2O5}. Thermo-optical properties are nonetheless greater than those of other m-OSR found in the literature so far.\cite{AZO,Ta2O5}

Although the Reststrahlen band of SiO$_2$ limits performance in the 8–11 $\mu$m range, the present optimization mitigates this constraint. Experimental validation of the proposed design constitutes a natural continuation of this work.

These results highlight the potential of metasurface-based optical solar reflectors for advanced thermal management in space and terrestrial radiative cooling applications.

\begin{acknowledgments}
The authors acknowledge K. Fleury-Frenette and L. Jacques from Centre Spatial de Liège (CSL) for stimulating discussions, as well as L. Weber from University of Namur for interesting brainstorming and insights validation with COMSOL. M. Lobet, and A. Mayer are
research associates of the Fonds de la Recherche Scientifique – FNRS. This research used resources of the "Plateforme
Technologique de Calcul Intensif (PTCI)" located at the University of Namur, Belgium, which is supported by the FNRS-FRFC,
the Walloon Region, and the University of Namur (Conventions No.2.5020.11, GEQ U.G006.15, 1610468, RW/GEQ2016 et
U.G011.22).
\end{acknowledgments}

\section*{Data Availability Statement}


\begin{table}[h]
\begin{ruledtabular}
\begin{tabular}{ll}
\textbf{AVAILABILITY} & \textbf{STATEMENT OF DATA} \\ 
\textbf{OF DATA} & \textbf{AVAILABILITY} \\ 
\hline
Data available on&
The data that support the findings of\\
request from the&
this study are available from the\\
authors &
corresponding author upon reasonable\\
&
request. \\
\end{tabular}
\end{ruledtabular}
\end{table}

\begin{figure*}
\includegraphics[width=1\textwidth,trim={0cm 10cm 0cm 2.5cm},clip]{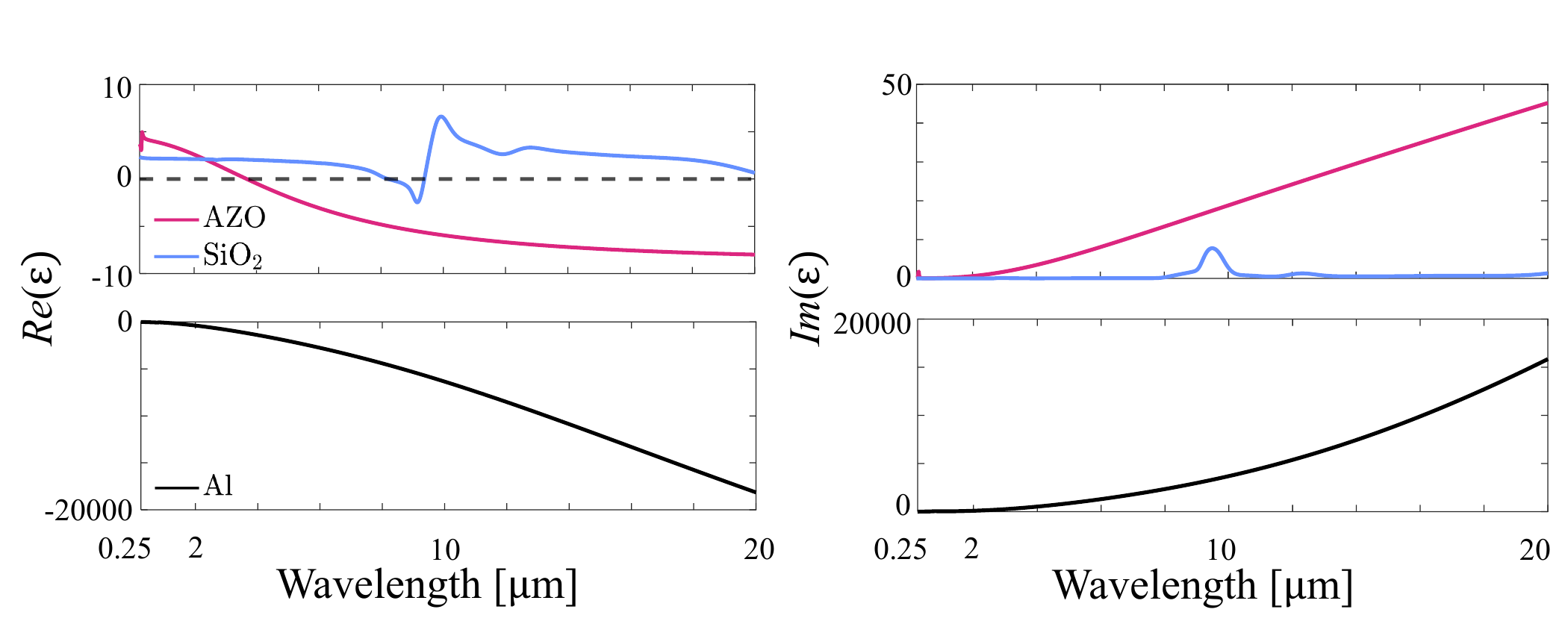}
\caption{\label{fig:7}Real and imaginary parts of the permittivity of the three materials used in the m-OSR design, AZO, SiO$_2$ and Al, between 0.25 and 20 microns.\cite{AZO,Franta,Rakic}}
\end{figure*}

\appendix

\section{Materials permittivity}\label{appendixes}

The permittivity of the materials used in the architecture of the m-OSR design is provided in \autoref{fig:7} between 0.25 and 20 $\mu$m. Al and AZO present a classical Drude model behavior,\cite{Rakic,AZO} with the exception of the latter UV absorption peak around 300 nm.\cite{AZO} SiO$_2$ exhibits an emission peak at 9 microns and oscillations of the real part of its permittivity, leading to the formation of a Reststrahlen band around 10 microns.\cite{Franta}

\section{Models parameters}

\autoref{tab:model} gives the fixed parameters used in the Drude-Lorentz model representing the permittivity of AZO (derived from Sun \textit{et al.}\cite{AZO}). The Drude model used for Al is given in Rakic \textit{et al.}\cite{Rakic}.
\begin{table}[h]
\caption{Parameters of the Drude-Lorentz model representing AZO permittivity.}
\begin{ruledtabular}
\begin{tabular}{lcr}
Parameter&Value&Units\\
\hline
$\varepsilon_{\infty}$ & 4.0&-\\
$\Gamma$ & 3.49$\cdot 10^{14}$&rad/s\\
$m^{*}/m_0$ & 0.4&-\\
$f_1$ & 0.1649&-\\
$\Gamma_1$ & 5.06$\cdot 10^{14}$&rad/s\\
$\omega_1$ & 5.96$\cdot 10^{15}$&rad/s\\
\end{tabular}
\end{ruledtabular}
\label{tab:model}
\end{table}

\newpage


\providecommand{\noopsort}[1]{}\providecommand{\singleletter}[1]{#1}%

\end{document}